# Tunable Light Emission from GaAsP and GaInP Islands Grown on Silicon (001) Nanotips Wafer

Navid Kafi,[1] Adriana Rodrigues,[1] Ines Häusler,[1] Haoran Ma,[1,2] Carsten Netzel,[3] Adnan Hammud,[4] Oliver Skibitzki,[5] Martin Schmidbauer,[2] and Fariba Hatami[1]

[1] *Institut für Physik, Humboldt-Universität zu Berlin, Newtonstr.15, 12489 Berlin, Germany*
[2] *Leibniz Institut für Kristallzüchtung, Max-Born Str.2, 12489 Berlin, Germany*
[3] *Ferdinand-Braun-Institut gGmbH, Leibniz-Institut für Höchstfrequenztechnik, Gustav-Kirchhoff-Str. 4, 12489 Berlin, Germany*
[4] *Department of Inorganic Chemistry, Fritz-Haber-Institute der Max-Planck-Gesellschaft, Faradayweg 4–6, 14195 Berlin, Germany*
[5] *IHP-Leibniz Institut für Innovative Mikroelektronik, Im Technologiepark 25, 15236 Frankfurt (Oder), Germany*

We present the monolithic integration of $GaAs_xP_{1-x}$ and $Ga_xIn_{1-x}P$ islands, selectively grown on a Si(001) nanotip wafer using gas-source molecular-beam epitaxy via a nanoheteroepitaxy approach. Optimal growth temperatures balancing selectivity and compositional control are 520–580 °C for GaAsP and 500–510 °C for GaInP. By adjusting the group III and V fluxes, island luminescence is tuned across a broad spectral range from 1.5 eV to 2.1 eV. High-resolution X-ray diffraction measurements on ensembles exceeding one million islands confirm that both GaAsP and GaInP islands are relaxed, while broad diffraction linewidths point to alloy fluctuations. Scanning transmission electron microscopy combined with energy-dispersive X-ray spectroscopy reveals compositional variations of up to 6% among the GaAsP islands, with nearly uniform composition within individual islands. In the case of GaInP, indium-rich regions are observed within single islands, with up to 11% variation across the ensemble. These compositional variations result in broadened or multiple luminescence peaks. Despite challenges in achieving full uniformity, this work demonstrates a pathway toward scalable III–V infrared-to-visible light emitters monolithically integrated on silicon, advancing microscale light sources and detectors for silicon photonics.

Silicon (Si) is the dominant platform for modern microelectronics, offering a mature, scalable, and cost-effective technology base. However, due to its indirect bandgap, Si is inherently inefficient at light emission—posing a major challenge for the realization of fully integrated silicon photonic systems. Various Si-based strategies have been explored to overcome this limitation, including alloying, the incorporation of hetero- and quantum structures, and the development of CMOS-compatible silicon avalanche-mode light-emitting diodes.[1-3] While these approaches offer promising solutions, the monolithic integration of III–V semiconductors onto Si has emerged as a particularly powerful alternative. This approach combines the scalability of Si technology with the superior optoelectronic properties of III–V materials, such as direct bandgaps and high carrier mobilities. Integrating III–V materials with mature silicon platforms opens pathways to scalable, cost-effective, and compact optoelectronic devices for next-generation silicon photonics.[4-8]

In particular, ternary III–V compounds such as $GaAs_xP_{1-x}$ and $Ga_xIn_{1-x}P$ are of significant interest due to their composition-tunable electronic structures and bandgaps, which span the near-infrared to visible spectrum. These materials undergo a composition-dependent direct-to-indirect bandgap transition: $GaAs_xP_{1-x}$ becomes indirect when the phosphorus concentration exceeds ~47%, and $Ga_xIn_{1-x}P$ transitions at gallium concentrations above ~70%.[9] This spectral tunability, along with the ability to engineer the nature of the bandgap, enables a wide range of photonic and optoelectronic applications. However, similar to binary III–V semiconductors, the monolithic integration of GaAsP and GaInP on Si remains technologically challenging due to significant material mismatches—including lattice constants, thermal expansion coefficients, and the polar–non-polar interface between III–V materials and Si. These mismatches lead to the formation of structural defects such as misfit dislocations, stacking faults, microtwins, and antiphase domains, all of which degrade device performance.[10]

Various methods have been developed to reduce defect formation during monolithic integration of III-V semiconductors on Si, including SiGe buffer layers as virtual substrates,[10-11] and mask-based methods such as selective area growth (SAG), and aspect ratio trapping (ART).[5-8,10,12-13] In mask-based methods, III-V semiconductors grow only at defined sites on patterned Si structures. In a related method, nanoheteroepitaxy (NHE), III-V semiconductors grow on pre-patterned nanoscale Si nanotips (NTs) embedded in $SiO_2$. Unlike the above, NHE manages strain through three-dimensional relaxation, using the compliance of Si nanotips.[14] Due to the size of the tips, the strain is partitioned between the epitaxial structures and the nanotips, decaying exponentially from the interface into both materials. Using NHE, several binary III-V semiconductors, such as, InP,[15,16] GaP,[17,18] GaAs,[19–20] and GaN [21] have grown as islands on Si NTs.

In this paper, we present the selective growth and properties of ternary semiconductor alloys, $GaAs_xP_{1-x}$ and $Ga_xIn_{1-x}P$ using the NHE approach. These materials were monolithically grown on Si(001) NTs using gas-source molecular-beam epitaxy (GS-MBE). We address the critical challenge of controlling alloy miscibility—and thereby composition—while ensuring growth selectivity on the nanotips, which is essential for precise spatial positioning. Our results demonstrate a class of broadband light emitters integrated with Si, spanning a spectral range from 1.5 to 2.1 eV. This advancement establishes a versatile platform for silicon photonics, in particular, for tunable micro-scale light sources and detectors.

Nanotip-patterned Si(001) substrates were fabricated using a state-of-the-art 0.13 μm SiGe BiCMOS technology pilot line on 200 mm wafers. The tips were arranged in square arrays with tip-to-tip distances ranging from 0.5 μm to 2 μm. Prior to growth, all substrates were cleaned in Piranha solution and etched in hydrofluoric acid (HF) to remove organic residues and the native $SiO_2$ layer. Details on the wafer fabrication and preparation process can be found elsewhere.[16,20]

After etching, the substrates were then immediately transferred into a Riber 21C gas-source MBE system. Before initializing growth, the substrates were heated to 720 °C in the growth chamber to remove any residual oxides from the nanotips that may have formed during the transfer. The temperature was then lowered to the desired growth temperature, after which epitaxy was initiated. The structures were grown using solid-source indium and gallium, and thermally cracked phosphine ($PH_3$) and arsine ($AsH_3$).

The surface morphology was characterized using a Raith Pioneer Two, Raith scanning electron microscopy (SEM) to primarily assess whether growth occurred selectively on the nanotips. To analyze the average alloy composition, high-resolution X-ray diffraction (HRXRD) measurements were performed using a 9 kW Rigaku SmartLab system with Cu K$\alpha_1$ radiation ($\lambda$ = 1.54056 Å). The optical properties were investigated using photoluminescence (PL) and cathodoluminescence (CL) spectroscopy. PL were measured over a temperature range of 6–300 K. CL measurements were conducted at 83 K using an Ultra Plus SEM (Zeiss) equipped with a MonoCL4 system (Gatan). Panchromatic CL images were captured with a photomultiplier, while spectral data were acquired using a Peltier-cooled CCD detector. The individual islands and their material composition were analyzed using scanning transmission electron microscopy (STEM) in combination with energy-dispersive X-ray spectroscopy (EDXS). More details can be found elsewhere.[18]

To achieve selective growth, the substrate temperature plays a critical role. Figure 1(a) summarizes the optimized substrate temperatures for the GS-MBE NHE of GaAs, GaP, and InP. Growth temperatures below the range lead to parasitic growth, while temperatures above this range result in reduced growth rates.[16-19] The insets in Fig. 1(a) present the top-view SEM images of GaP islands, demonstrating both parasitic and selective growth, and the cross-section TEM image of a selectively grown GaP island on a Si nanotip. The overlap of the optimized substrate temperature ranges for two binary semiconductors approximately defines the range suitable for the selective growth of their corresponding alloy. These overlapping temperature windows were used as starting points for the nanoheteroepitaxy of $Ga_xAs_{1-x}P$ and $Ga_xIn_{1-x}P$ marked by the red and green squares in Fig. 1(a).

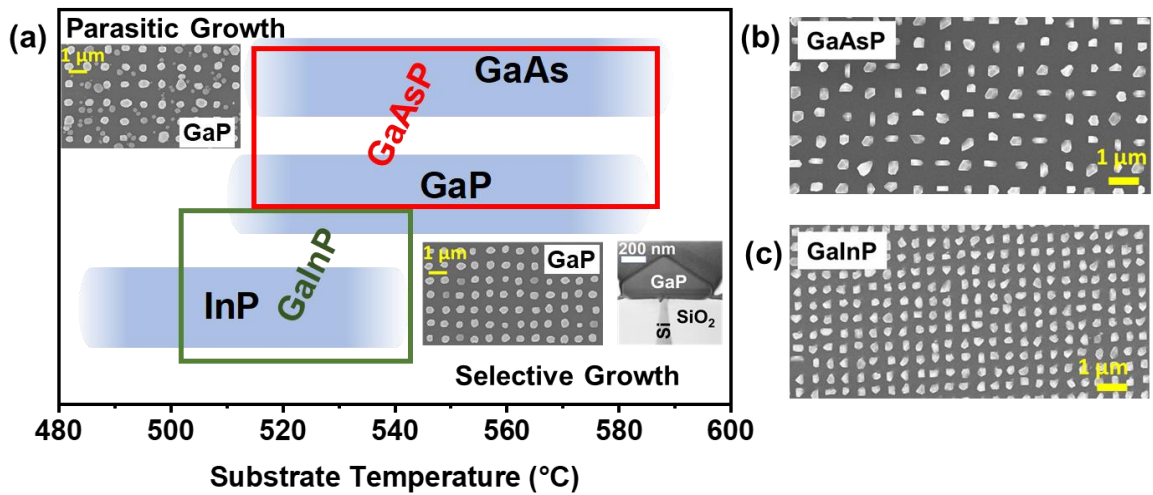

**FIG. 1.** Overview of nanoheteroepitaxy of III-V semiconductors. **(a)** Summary of the optimized substrate temperature range (highlighted in blue) for NHE of GaAs, GaP, and InP. The insets show the top-view SEM images of two GaP samples (parasitic and selective growth) and the CS-TEM image of a single GaP island. The red and green highlighted squares indicate the expected temperature ranges for the selective growth of GaAsP and GaInP, respectively. **(b), (c)** Top-view SEM images of selectively grown GaAsP and GaInP islands.

The temperature ranges for the nanoheteroepitaxy of GaAs and GaP overlap by about 60 °C, while GaP and InP exhibit a narrower overlap of 40 °C. This difference is attributed to the different vapor pressures of indium and gallium, a factor that influences the difference in the growth temperatures of the epitaxial InP and GaP. As a result, the window of the nanoheteroepitaxy for GaInP is narrower than that for GaAsP. Fig. 1(b) and (c) present top-view SEM images of selectively grown GaAsP and GaInP islands. In the following, we first discuss the NHE conditions and properties of GaAsP, an alloy with a narrower miscibility gap compared to GaInP.[22]

The GaAsP samples were grown at 565 °C with a Ga rate of $R_{Ga} = 0.3$ Å/s but different flux of phosphine ($F_{PH_3}$) and arsine ($F_{AsH}$). As shown in Fig. 1(b), smaller islands are rectangular, while larger ones exhibit multiple facets. Size analysis reveals a bimodal distribution. Figure 2(a) shows HRXRD 2θ-ω-scans of two samples, each containing more than $10^6$ islands. Sample #1 was grown with $F_{PH_3} = F_{AsH_3} = 1$ sccm ($F_{PH_3}/F_{AsH_3} = 1$), while Sample#2 was grown with $F_{PH_3} = 1.2$ sccm, $F_{AsH_3} = 1$ sccm ($F_{PH_3}/F_{AsH_3} = 1.2$). Each scan exhibits a sharp, intense peak at 2θ = 69.13°, corresponding to the 004 Bragg reflection of the Si substrate. In addition, broad Bragg reflections are observed at 2θ = 67.19° (Sample #1) and 2θ = 67.55° (Sample #2), arising from the $GaAs_xP_{1-x}$ islands. X-ray reciprocal space maps indicate that the islands are fully relaxed in both samples. Vegard's rule was therefore applied to estimate the average alloy composition across the ensemble of islands, yielding arsenic atomic fractions of x = 0.58, and x = 0.45 for Samples #1 and #2, respectively. The expected Bragg reflection positions for these compositions are indicated by the dashed lines in Fig. 2(a). Note that $GaAs_{0.58}P_{0.42}$ is expected to have a direct bandgap, while $GaAs_{0.45}P_{0.55}$ exhibits an indirect bandgap.

Given the mass flow controller tolerance of ± 0.1 sccm and the $AsH_3$ higher decomposition efficiency over $PH_3$,[23] the measured compositions align well with the targets, confirming controllable GaAsP alloy tuning via $PH_3$ and $AsH_3$ fluxes. For the sake of clarity, we will refer to Samples #1 and #2 as $GaAs_{0.58}P_{0.42}$ and $GaAs_{0.58}P_{0.42}$, respectively.

Figure 2(b) presents the photoluminescence spectra (8 K) for both samples, excited by a 440 nm laser diode (P = 45 mW). The signal originates from several hundred thousand islands. The spectra exhibit multiple peaks with overlap between 1.9 and 2.0 eV. The PL of $GaAs_{0.58}P_{0.42}$ sample peaks at 1.95 eV with a shoulder at 1.91 eV, and the spectrum of $GaAs_{0.45}P_{0.55}$ sample peaks at 2 eV accompanied by a shoulder at 2.05 eV. The higher PL energy of $GaAs_{0.45}P_{0.55}$ can be explained by 13% higher P content, which increases the bandgap. The PL of $GaAs_{0.58}P_{0.42}$ is detectable up to 300 K (Fig. 2(b)), whereas that of $GaAs_{0.45}P_{0.55}$ is quenched more rapidly with increasing temperature. This trend is consistent with the average alloy compositions, suggesting a higher proportion of direct bandgap islands in $GaAs_{0.58}P_{0.42}$ sample than in $GaAs_{0.45}P_{0.55}$ sample. The temperature evolution of the $GaAs_{0.58}P_{0.42}$ PL peak energy is well reproduced by the Varshni equation (Fig. 2(c)).[24]

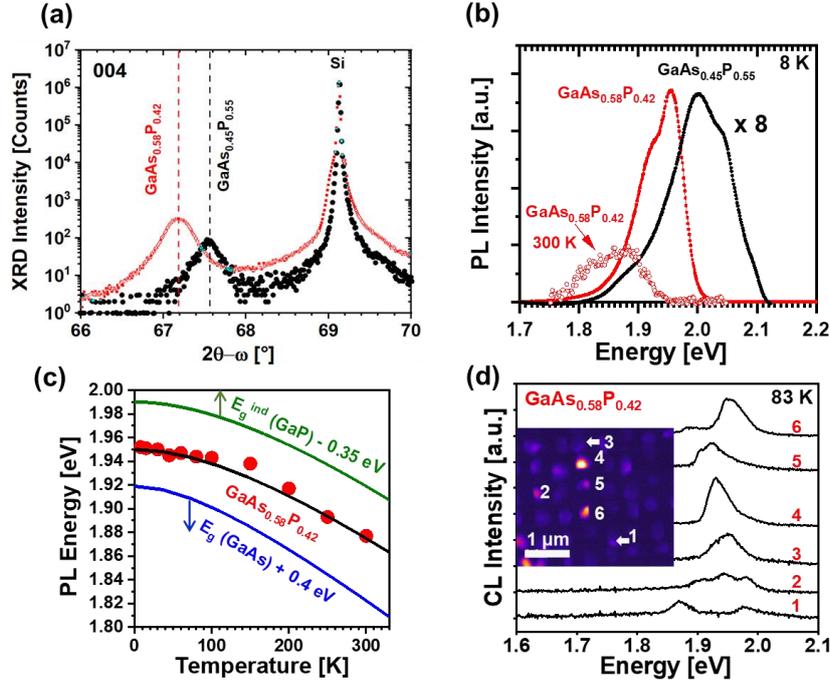

**FIG. 2.** Properties of the GaAs$_{0.58}$P$_{0.42}$ and GaAs$_{0.45}$P$_{0.55}$ samples. HRXRD 2θ-ω-scans **(a)** and PL spectra **(b)** collected from large ensembles of islands. The vertical dashed lines mark the expected Bragg position for GaAsP with As contents of 45% and 58%. **(c)** The temperature dependence shift of the high energy PL peak of GaAs$_{0.58}$P$_{0.42}$ sample and the corresponding fit using Varshni eq. (α=5.7×10$^{-4}$ eV/K, β=390 K). The expected shift for the relaxed GaP and GaAs are given with offset for comparison. **(d)** Micro CL spectra of several GaAs$_{0.58}$P$_{0.42}$ islands.

Figure 2(d) presents the CL intensity image of the GaAs$_{0.58}$P$_{0.42}$ sample and selected CL spectra from six islands. Several islands exhibit a dominant CL peak, varying in energy between 1.83 and 1.95 eV. Since the islands are epitaxially grown and relaxed, variations in emission likely originate from differences in alloy composition. Furthermore, defects, surface/interface states, and local disorder can critically affect luminescence. A comprehensive investigation of these effects requires correlated structural and luminescence measurements on the same individual island. Note, the islands are too large for quantum effects, and for emission broadness arising from size distribution.

We now turn our discussion to the NHE of GaInP. Direct band gap GaInP alloys cover a broader spectral range (1.35–2.2 eV) than direct bandgap GaAsP (1.42–2.1 eV),[9,25] making GaInP more advantageous for light-emitting devices at shorter and longer wavelength. GaInP samples were grown using indium rates, $R_{In}$ = 0.3–0.9 Å/s, and $R_{Ga}$= 0.2–0.4 Å/s (1.5> $R_{In}$/$R_{Ga}$ >4.5), and a PH$_3$ flux of 2.3 sccm.

Higher growth temperatures are expected to enhance miscibility and prevent phase separation. However, elevated temperatures also increase the indium desorption rate, which reduces indium incorporation and leads to Ga-rich alloys. To address this trade-off, we investigated the selective growth temperature window (green square in Fig. 1(a)) to identify the optimal conditions. A top-view SEM image of a selectively grown GaInP sample is shown in Fig. 1(c).

Figure 3(a) presents HRXRD 2θ-ω-scans of four samples differing in growth temperature $T_{sub}$, and $R_{In}/R_{Ga}$ ratio. Since the Si and GaInP Bragg peaks appear in close proximity we have chosen the 002 Bragg reflection for which the signal from the Si substrate at 2θ = 32.96° is suppressed and can be clearly distinguished from the broad GaInP 002 Bragg reflection. For the two high-temperature samples ($T_{sub}$ ≥ 525 °C) the GaInP island peaks are located around 2θ = 32.6° (blue and red curves in Fig. 3(a)). This corresponds to an average Ga content of x = 0.92 ± 0.02, resulting in an indirect-bandgap alloy. This value is significantly higher than the targeted Ga content, due the high indium desorption rate and remains $R_{In}/R_{Ga}$ ratio independent within this temperature range.

By reducing the growth temperature, indium desorption can be suppressed, which helps to maintain the intended indium content in the islands. The HRXRD scans of two samples grown at 510 °C and 500 °C with a ratio of $R_{In}/R_{Ga}$ = 1.5 (black and green curves, in Fig. 3(a)), exhibit distinct 002 GaInP peaks at 2θ = 31.83° and 2θ = 31.35°, corresponding to the averaged alloy compositions of $Ga_{0.6}In_{0.4}P$ and $Ga_{0.4}In_{0.6}P$, respectively. This confirms enhanced indium incorporation and the achievement of the targeted alloy composition, $Ga_{0.4}In_{0.6}P$, for the sample grown at 500 °C. However, the broadness of the 002 GaInP peak suggests a distribution in alloy composition across the ensemble, and/or contributions from defects. Growth temperatures below 500 °C results in parasitic growth on the $SiO_2$ mask, which impairs selectivity. Consequently, for the fluxes used here, the optimal growth window is narrowed to 500–510 °C, balancing both selectivity and compositional control. A lower flux may reduce parasitic growth and widen the optimal temperature range.

To assess the light emission properties of the GaInP islands, their PL and CL spectra were studied. For our discussion, we selected two representative samples: $Ga_{0.6}In_{0.4}P$ with an average direct-bandgap composition, and $Ga_{0.92}In_{0.08}P$ with an indirect bandgap. Figure 3(b) presents the PL spectra of both samples. The $Ga_{0.6}In_{0.4}P$ sample exhibits a broad PL signal (1.5–2.1 eV) centered at 1.7 eV, persisting up to 300 K. The $Ga_{0.92}In_{0.08}P$ sample shows a weaker but narrower PL peak at 2.11 eV, detectable only up to 80 K, consistent with phonon-assisted excitonic recombination in indirect semiconductors.[17]

The PL peak energies of both samples are lower than the calculated bandgap energies based on the estimated alloy composition using the XRD data in Fig. 3(a). This discrepancy arises because XRD provides an average alloy composition across an ensemble of over one million islands, whereas PL reflects radiative recombination processes. Carriers tend to migrate toward the localized regions with lower bandgap energy, resulting in a shift of the PL emission to lower energies. Such low-bandgap regions may originate from indium-rich segments caused by alloy composition fluctuations.

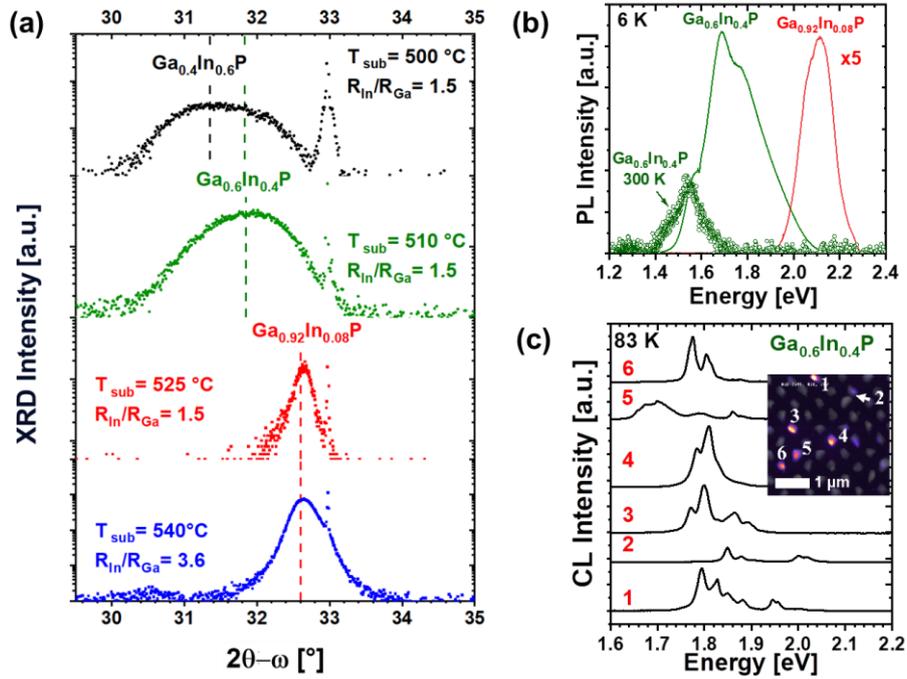

**FIG. 3.** Properties of GaInP islands. **(a)** HRXRD 2θ-ω-scans for samples grown at varying temperatures and $R_{In}/R_{Ga}$ ratios. The vertical dashed lines mark the expected Bragg position for GaInP alloys with Ga contents of 92%, 60% and 40%. **(b)** PL spectra of $Ga_{0.6}In_{0.4}P$ (6 K and 300 K) and $Ga_{0.92}In_{0.08}P$ sample (6 K). **(c)** Micro-CL spectra of multiple $Ga_{0.6}In_{0.4}P$ islands.

Figure 3(c) presents the CL intensity image of the $Ga_{0.6}In_{0.4}P$ sample along with the spectra from six individual islands. The islands exhibit variations in emission behavior, with multiple emission lines ranging from 1.6 to 2.05 eV. This multiplicity may be related to compositional variations within individual islands, where localized material differences lead to different bandgap energies.

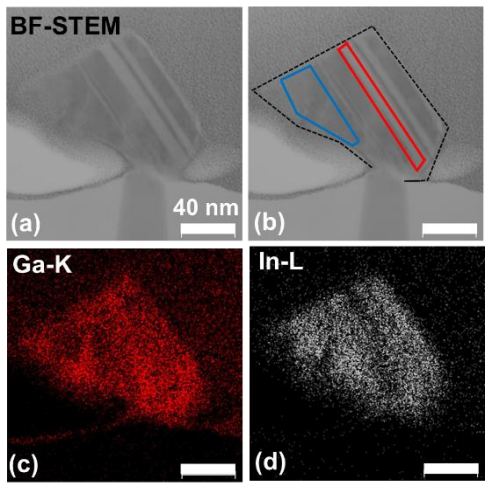

**FIG. 4.** **(a), (b)** Bright-field STEM image of a $Ga_{0.6}In_{0.4}P$ island and **(c), (d)** corresponding 2D EDXS elemental maps. The specific X-ray absorption edges used for identifying Ga and In (K and L edges) are indicated above each map. The region enclosed by blue lines in the STEM image (b) exhibits an In concentration approximately 11% higher (relative) than that of the red-lined region, based on quantitative EDX analysis. The scale bar in all images corresponds to 40 nm.

To gain deeper insight into the material composition of single islands, several islands were analyzed using STEM in combination with EDXS. The data indicate that for the $GaAs_{0.58}P_{0.42}$ sample the alloy composition varies by about 6% among the islands, while each island itself remains nearly homogeneous. In contrast, the GaInP islands show compositional variations both among and within the islands, with Ga- and In-rich segments coexisting. In contrast, the GaInP islands exhibit compositional variations both between and within individual islands, with Ga-rich and In-rich segments coexisting. Figure 4(a) shows a bright-field STEM image of a representative $Ga_{0.6}In_{0.4}P$ island, where several regions with varying contrast are visible. Two distinct regions, highlighted in red and blue in Fig. 4(b), were further analyzed. Elemental maps based on the Ga-K and In-L X-ray absorption edges, shown in Fig. 4(c) and (d), respectively, reveal spatial differences in Ga and In distribution within the island. According to the EDXS analysis, the area outlined in blue contains 11% more indium than the red-marked region, indicating notable intra-island compositional variation. These findings are consistent with the CL data in Fig. 2(d) and 3(c), which show that most $GaAs_{0.58}P_{0.42}$ islands exhibit a dominant emission peak that varies from island to island, whereas individual $Ga_{0.6}In_{0.4}P$ islands exhibit multiple CL peaks.

The difference in compositional fluctuations between GaInP and GaAsP islands can be understood from a thermodynamic perspective, where the driving force for phase separation and the formation of a miscibility gap are determined by the enthalpy of mixing ($\Delta H_{mix}$).[22] The interaction between GaP and InP leads to a significantly higher positive $\Delta H_{mix}$ compared to that between GaP and GaAs, resulting in a stronger tendency for compositional fluctuations in GaInP than in GaAsP.[26] Higher growth temperatures enhance atomic diffusion and entropy of mixing, thus promoting better alloy homogeneity. However, the high vapor pressure of indium imposes a practical limit on the maximum growth temperature for GaInP islands, as supported by HRXRD analysis in Fig. 3(a).

It is now worth returning to the CL intensity images (insets of Fig. 2(d) and 3(c)), which show that not all islands in the $GaAs_{0.58}P_{0.42}$ and $Ga_{0.6}In_{0.4}P$ samples emit light. The compositional fluctuations discussed above can explain this behavior. A local increase of 5% in P or 10% in Ga content in the $GaAs_{0.58}P_{0.42}$ and $Ga_{0.6}In_{0.4}P$ alloys, respectively, can shift their band structure from direct to indirect. Such a shift determines whether an individual island can function as an efficient light emitter. Islands with an indirect bandgap—where non-radiative recombination dominates and luminescence occurs primarily via phonon-assisted excitonic recombination—appear "dark" at the elevated temperature of 83 K, as observed in the temperature-dependent PL spectra. In addition, due to the small size of the islands, carriers can rapidly reach the surface and recombine non-radiatively through surface states. Surface passivation and improved compositional control may therefore enhance the emission efficiency of individual islands and, consequently, the overall light output efficiency.

In conclusion, we have demonstrated the monolithic integration of $GaAs_xP_{1-x}$ and $Ga_xIn_{1-x}P$ islands via NHE by gas-source MBE on Si(001) nanotip wafers fabricated through a CMOS-compatible process flow. The optimal growth temperatures to balance both selectivity and miscibility are 520– 580 °C for GaAsP and 500– 510 °C for GaInP. The island compositions were tuned by adjusting group III and V fluxes, enabling luminescence energies from 1.5 to 2.1 eV. HRXRD measurements of ensembles confirm that the islands are relaxed; however, the broad linewidths also indicate variations in alloy composition. Spatially resolved STEM-EDXS analysis reveals compositional inhomogeneities among the islands up to 11%. The individual GaAsP islands appear compositionally uniform, while in GaInP, In-rich segments are observed within single islands. These fluctuations manifest in the luminescence properties as broadened or multiple emission peaks. Challenges remain in achieving precise alloy homogeneity and in improving surface passivation to enhance emission efficiency. Nevertheless, this work lays the foundation for scalable, broadband III-V light emitters on Si, advancing integrated photonic applications.

Acknowledgments: This work was supported by German Research Foundation (DFG- Grant No: 428250328). We also thank the European Regional Development Fund ERDF (project number 1.8/15).